\documentclass[12pt]{iopart}
\usepackage{iopams}

\usepackage{graphicx}
\usepackage{color}
\usepackage{bm}

\newcommand{\eqname}[1]{\label{eq:#1}}

\newcommand{\eq}[1]{(\ref{eq:#1})}
\newcommand{\eqs}[2]{(\ref{eq:#1}-\ref{eq:#2})}

\newcommand{\Psihd}{\hat\Psi^\dagger}

\newcommand{\Psih}{\hat\Psi}

\newcommand{\ahd}{\hat a^\dagger}
\newcommand{\ah}{\hat a}

\newcommand{\bh}{\hat b}
\newcommand{\bhd}{\hat{b}^\dagger}

\begin{document}

\title[Acoustic white holes in atomic BECs]
{Acoustic white holes in flowing atomic Bose-Einstein condensates}

\author{Carlos Mayoral$^{1}$, Alessio Recati$^{2}$, Alessandro Fabbri$^{1,3}$, Renaud Parentani$^{4}$,
 Roberto Balbinot$^5$,  Iacopo Carusotto$^{2}$
}

\address{$^1$ Departamento de F\'isica Te\'orica and IFIC, Universidad de Valencia-CSIC,
C. Dr. Moliner 50, 46100 Burjassot, Spain}
\address{$^2$ INO-CNR BEC Center and Dipartimento di Fisica,
Universit\`a di Trento,  \\ via Sommarive 14, I-38123 Povo, Italy}
\address{$^3$ APC (Astroparticules et Cosmologie), 10 rue A. Domon et L. Duquet, 75205 Paris Cedex 13, France}
\address{$^4$ Laboratoire de Physique Th\'eorique, CNRS UMR 8627, B\^at. 210, Universit\'e
Paris-Sud 11, 91405 Orsay Cedex, France}
\address{$^5$ Dipartimento di Fisica dell'Universit\`a di Bologna and
INFN sezione di Bologna, Via Irnerio 46, 40126 Bologna, Italy}

\ead{carusott@science.unitn.it}

\begin{abstract}
We study acoustic white holes in a steadily flowing atomic Bose-Einstein condensate. A white hole configuration is obtained when the flow velocity goes from a super-sonic value in the upstream region to a sub-sonic one in the downstream region.
The scattering of phonon wavepackets on a white hole horizon is numerically studied in terms of the Gross-Pitaevskii equation of mean-field theory: dynamical stability of the acoustic white hole is found, as well as a signature of a nonlinear back-action of the incident phonon wavepacket onto the horizon.
The correlation pattern of density fluctuations is numerically studied by means of the truncated-Wigner method which includes quantum fluctuations. Signatures of the white hole radiation of correlated phonon pairs by the horizon are characterized; analogies and differences with Hawking radiation from acoustic black holes are discussed. In particular, a short wavelength feature is identified in the density correlation function, whose amplitude steadily grows in time since the formation of the horizon. The numerical observations are quantitatively interpreted by means of an analytical Bogoliubov theory of quantum fluctuations for a white hole configuration within the step-like horizon approximation.
\end{abstract}

\date{\today}
\submitto{\NJP}


\section{Introduction}

The recent experimental advances in the creation and manipulation of atomic Bose-Einstein condensates are suggesting these systems as ideal candidates where to experimentally address fundamental questions in quantum hydrodynamics~\cite{BECbook}. In particular, a lot of attention has been devoted to configurations showing regions of super-sonic flow, where the dynamics is the richest and quantum effects should be easiest to observe.
Among the most remarkable recent observations, we may mention the emission of a wake of Bogoliubov phonons by a moving defect via an analog of the \u Cerenkov effect~\cite{Cerenkov}, the shedding of solitons in a one-dimensional configuration~\cite{engels}, the onset of dynamical instabilities in the periodic potential of optical lattices~\cite{FI}. In the meanwhile, theoretical investigations have addressed general questions about the stability of superflow~\cite{wuniu} and have identified the mechanisms underlying the shedding of solitons~\cite{hakim} and vortices~\cite{jackson}. Analytical aspects of the one-dimensional flow have been also investigated in detail~\cite{kamchatnov}. Most of this physics is accurately described within a mean-field approach, where the dynamics of the Bose-condensed atomic cloud is described in terms of the Gross-Pitaevskii equation for the condensate wavefunction~\cite{BECbook,castin}.

In the wake of the formal analogy~\cite{unruh,analogy_book} between the propagation of quantum fields on a curved space-time background~\cite{c-QFT} and the propagation of sound on inhomogeneously moving fluids, a strong interest is presently being devoted to the quantum features of the condensate hydrodynamics.
A most intriguing prediction concerns the emission of correlated pairs of phonons by the horizon of an acoustic black hole configuration via a mechanism which is a condensed-matter analog of the well-celebrated Hawking radiation from gravitational black holes~\cite{hawking}. Following the proposal of~\cite{PRA_analytic}, the key signature of this effect was identified in the correlation function of density fluctuations: the correlation between the Hawking phonon and the partner falling into the black hole results in a peculiar, tongue-shaped feature in the correlation pattern. This prediction was confirmed in an {\em ab initio} numerical simulation of the dynamics of an atomic BEC~\cite{NJP_BH}. A first experimental attempt to create an acoustic black hole in an atomic condensate was reported in~\cite{steinhauer}.

As we schematically show in Fig.\ref{fig:sketch}, an {\em acoustic white hole} configuration is obtained by simply reversing the direction of flow of a black hole one: atoms go from an upstream, inner region where the flow is super-sonic to a downstream, external region where the flow is sub-sonic: within a naive hydrodynamic picture, dragging by the moving condensate forbids low-energy phonons from crossing the horizon and penetrating the inner region of the white hole.
Questions about the stability of such acoustic white hole configurations has lately attracted a lot of interest, and different answers have been proposed to this problem, ranging from instability~\cite{ulf}, to stability~\cite{macher2}, to intermediate behaviours depending on boundary conditions~\cite{barcelo}. Issues related to the quantum emission of phonons by configurations involving white hole horizons have been addressed in~\cite{macher2,macher1,coutant,finazzi}.

The present paper reports a comprehensive theoretical investigation of the physical properties of acoustic white holes in atomic Bose condensates. Focussing our attention on a specific, yet realistic configuration which is most suitable to numerical and analytical study, we address the problem of the dynamical stability of white hole configurations and then the physical properties of the {\em white hole radiation} emitted by the horizon. In order to be able to safely isolate the dynamics of the white hole horizon from spurious effects due to the finite size and/or the multiply connected geometry of the system~\cite{ring}, all numerical calculations have been performed using absorbing boundary conditions for the Bogoliubov phonons so to model the most relevant situation of an infinitely long condensate.

In Sec.\ref{sec:system} we introduce the physical system and we discuss the main features of the dispersion of Bogoliubov excitations in respectively the super- and sub-sonic regions: this will be the essential ingredient to physically understand the numerical observations that are presented in the following of the paper. Among the different possible choices of the density and flow velocity pattern that result in a white hole horizon, we concentrate our attention on a configuration that appears most robust against spurious effects. 

The dynamical stability of the white hole configuration is assessed in Sec.\ref{sec:stab}, where the scattering of a phonon wavepacket on the white hole horizon is studied by means of the Gross-Pitaevskii equation of mean-field theory. In addition to the reflected wavepacket in the outward direction, a pair of short wavelength wavepackets penetrate inside the white hole: as a consequence of the strong blue shift experienced while approaching the horizon, low energy hydrodynamic modes are in fact transformed into high-momentum single particle excitations that are able to enter the white hole. One of the two wavepackets consists of negative-norm Bogoliubov modes and results from the same mixing of positive and negative norm modes that underlies Hawking radiation.
Dynamical stability of the configuration is demonstrated by the fact that no exponentially growing deformation of the horizon develops under the effect of the incident wavepacket~\footnote{A similar procedure for more complex black hole/white hole configurations has confirmed the {\em black hole laser} phenomenon~\cite{coutant,bhlaser}, i.e. the onset of an exponentially growing perturbation of the density profile. This will be the subject of a forthcoming publication.}. The only visible back-action effect consists of the Bogoliubov-\u Cerenkov emission of zero-frequency phonons~\cite{Cerenkov} in the upstream direction by the slight deformation of the horizon; a specific discussion of this feature is given in Sec.\ref{sec:back}. As a function of time, the amplitude of this pattern eventually tends to a finite amplitude value, roughly proportional to the square of the incident wavepacket amplitude.

The effect of quantum fluctuations is studied in Sec.\ref{sec:quantumfluct} using the same truncated Wigner method as in~\cite{NJP_BH}. In analogy to what was done for acoustic black holes, our key observable is the correlation function of density fluctuations. Several characteristic features are identified and related to the correlated emission of phonon pairs by the white hole horizon into different pairs of modes. In spite of some similarities with the Hawking emission from black holes, the physical properties of the emission turn out to be significantly different: for this reason, we have chosen to refer to it under the different name of {\em white hole emission}. While Hawking radiation mostly consists of low-$k$ excitations, the phonons that are emitted into the white hole have a large momentum comparable to the healing length of the condensate; furthemore, the checkerboard pattern that appears in the correlation function for points located inside the white hole keeps growing in time with a characteristic logarithmic or linear law depending on the initial temperature. This is to be contrasted to the late-time stationary state of Hawking radiation from black holes.

An analytical understanding of the white hole radiation is developed in Sec.\ref{sec:analytic} based on the Bogoliubov theory of dilute Bose Einstein condensates within the step-like horizon approximation of~\cite{PRA_gradino}. Dynamical stability of the white hole configuration is assessed by checking that complex frequency eigenmodes of the Bogoliubov-de Gennes equations are indeed absent. A theory of quantum fluctuations is developed in terms of the $S$-matrix describing phonon scattering on the white hole horizon. This provides an analytical expression for the spectral distribution of the white hole emission: in contrast to the thermal character of Hawking radiation from black holes, the phonons that are emitted outside the white hole have a flat spectral distribution.
The same approach is then used to evaluate the different contributions to the correlation pattern: the resulting analytical expressions reproduce in a fairly accurate way all the features that are observed in the numerical calculations. The steady growth in time of the checkerboard pattern is related to an infra-red divergence of the density correlation function at late times: the logarithmic/linear growth is recovered by imposing a suitable infra-red cut-off to the integral.
Conclusions are finally drawn in Sec.\ref{sec:conclu}.

\section{The physical system and the theoretical model}
\label{sec:system}

\begin{figure}[htbp]
\begin{center}
\includegraphics[width=0.45\columnwidth,angle=0]{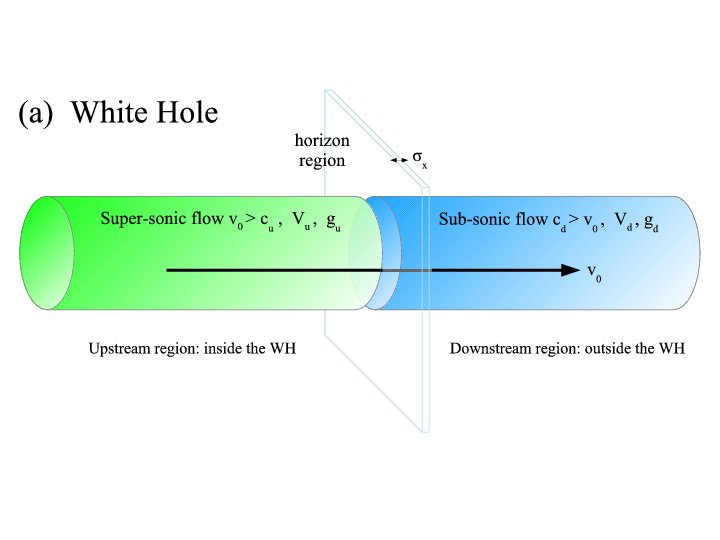}
\hspace*{0.05\columnwidth}
\includegraphics[width=0.45\columnwidth,angle=0]{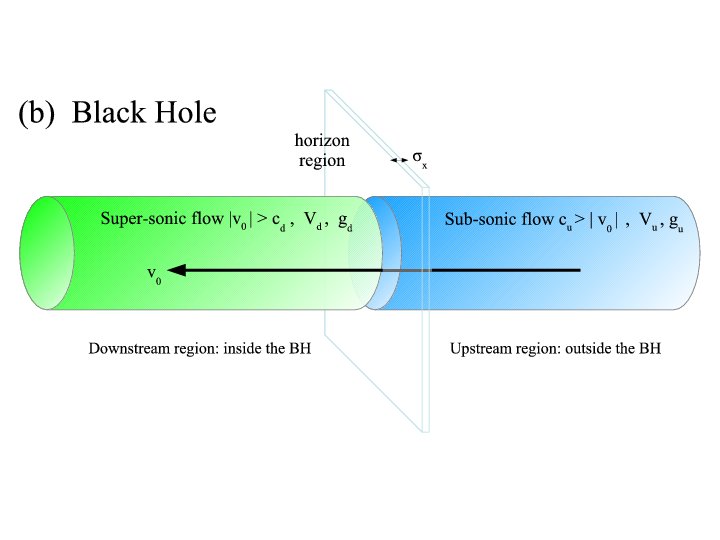}
\caption{Left panel: sketch of the acoustic white hole configuration.
Right panel: sketch of an acoustic black hole configuration.
The two configurations are related by a reversal of the condensate flow speed $v_0$.}
\label{fig:sketch}
\end{center}
\end{figure}

A sketch of the physical system we are considering is shown in the left panel of Fig.\ref{fig:sketch}. An elongated atomic Bose-Einstein condensate is steadily flowing along an atomic waveguide in the $\hat{x}$ direction. A suitable modulation of the confinement potential and/or of the atom-atom scattering length creates a {\em white hole} horizon at $x=0$ separating a upstream $x<0$ region of super-sonic flow from a downstream $x>0$ one of sub-sonic flow.
In the gravitational analogy~\cite{unruh,Rivista_2005,Barcelo-LIVING-REVIEW},
the upstream region corresponds to the interior of the white hole, while the downstream region corresponds to the external space.
For comparison, a sketch of the acoustic black hole configuration obtained reversing the flow speed is shown in the right panel~\footnote{Note that this configuration differs from the one considered in~\cite{NJP_BH,PRA_gradino} by a simple spatial inversion}.

To simplify the theoretical description, the transverse confinement is assumed to be tight enough for the transverse degrees of freedom to be frozen~\cite{BECbook} and the system dynamics to be accurately described by a one-dimensional model based on the following second-quantized Hamiltonian,
\begin{eqnarray}
\mathcal{H}&=&\int\!dx\,\Big[\frac{\hbar^2}{2m}\,\nabla \Psihd(x) \, \nabla\Psih(x)+V(x,t)\,\Psihd(x)\,\Psih(x)\,+ \nonumber \\
&+&\frac{g(x,t)}{2}\,\Psihd(x)\,\Psihd(x)\,\Psih(x)\,\Psih(x) \Big].
\end{eqnarray}
Here, $\Psih(x)$ and $\Psihd(x)$ are atomic field operators satisfying Bose commutation rules $[\Psih(x),\Psihd(x')]=\delta(x-x')$, $m$ is the atomic mass, $V(x)$ the external potential, and $g(x)$ is the effective one-dimensional atom-atom interaction constant~\cite{BECbook}.
At mean-field level, the condensate dynamics is described by the one-dimensional Gross-Pitaevskii equation (GPE)~\cite{BECbook}
\begin{equation}
i\hbar\,\frac{\partial \psi}{\partial t}=-\frac{\hbar^2}{2m}\,\frac{\partial^2 \psi}{\partial x^2}+V(x,t)\,\psi+g(x,t)\,|\psi|^2\,\psi
\eqname{GPE}
\end{equation}
for the macroscopic condensate wavefunction $\psi$.
In Sec.\ref{sec:stab}, this equation will be used to study the propagation of classical perturbations on top of the white hole configurations: full inclusion of the nonlinearity will allow us to study the dynamical stability of the configuration beyond the
linearized
Bogoliubov theory.
In the following Sec.\ref{sec:quantumfluct}, the so-called truncated Wigner method will be adopted to go beyond the mean-field approximation underlying \eq{GPE} and include the effect of quantum and thermal fluctuations: these are described in terms of stochastic noise on the initial value of the condensate wavefunction, which then evolves according to the same GPE \eq{GPE}. Expectation values of observables are obtained as averages over the stochastic noise.
We refer to~\cite{NJP_BH} for a detailed description of the numerical technique.

For the sake of simplicity, we shall focus our attention onto a specific configuration which is most suitable for both analytic and numeric investigation.
Initially, the condensate is assumed to have a spatially uniform density $n_0$ and a spatially uniform flow speed $v_0$ along the positive $x$ direction. The external potential and the (repulsive) atom-atom interaction constant are also uniform and equal to respectively $V(x)=V_d$ and $g(x)=g_d>0$.
Around $t=t_0$, a step-like spatial modulation is applied to both the potential and the interaction constant by suitably modifying the transverse confinement potential and/or the atom-atom scattering length via an external magnetic field tuned in the vicinity of a so-called Feshbach resonance~\cite{BECbook}: within a short time $\sigma_t$, $V$ and $g$ in the upstream $x<0$ region are brought to their final values $V_u$ and $g_u$, while their values in the $x>0$ region are kept equal to the initial ones $V_d$ and $g_d$. The transition region around $x=0$ has a characteristic thickness $\sigma_x$; away from it, both $V$ and $g$ quickly tend to their asymptotic values $V_{u,d}$ and $g_{u,d}$.

To avoid the development of spurious instabilities due to the reflection of Bogoliubov excitations at the edges of the integration box, suitable absorbing boundary conditions are implemented for the Bogoliubov modes. In practice, this is done by including a restoring force on the wavefunction that damps out Bogoliubov excitations and brings it back to a pure condensate. For this additional term not to interfere with the physics under investigation, it is spatially restricted to small regions at the edges of the integration box which are sufficiently far from the white hole horizon.

In order to reduce the impact of competing processes such as back-scattering of condensate atoms and soliton shedding from the potential step~\cite{hakim}, the external potential $V$ is chosen to exactly compensate the spatial jump in the Hartree interaction energy $\mu_{u,d}=g_{u,d}\,n_0$, i.e. 
\begin{equation}
V_d+\mu_d=V_u+\mu_u.
\eqname{Hartree}
\end{equation}
In this way, the plane wave
\begin{equation}
\psi(x,t)=\sqrt{n_0}\,\exp[i(k_0 x-\omega_0 t)]
\eqname{planewave}
\end{equation}
with $\hbar k_0/m =v_0$ and $\omega_0=\hbar k_0^2/2m$ is for all times a solution of the GPE \eq{GPE}.

\begin{figure}[htbp]
\begin{center}
\includegraphics[width=0.8\columnwidth,angle=0,clip]{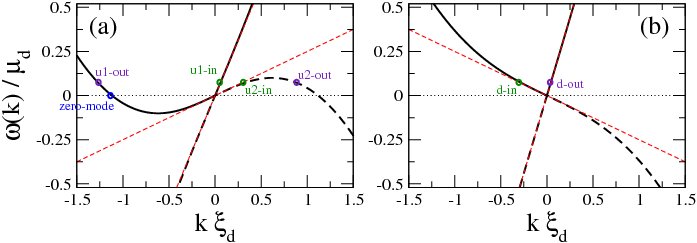}\\
\vspace{0.5cm}
\includegraphics[width=0.8\columnwidth,angle=0,clip]{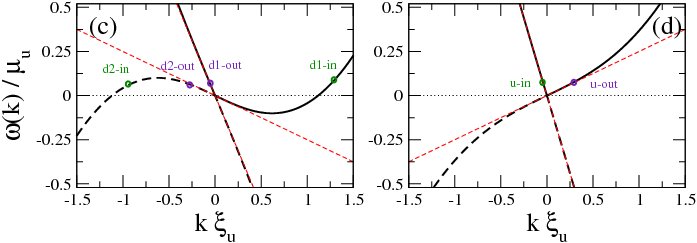}
\caption{
Dispersion of Bogoliubov excitations in the asymptotic regions far away from the horizon. The labels of the modes indicate the up- or down-stream region with respect to the condensate flow and the in- or out-going direction of their propagation direction (as predicted by their group velocity) with respect to the horizon~\footnotemark. Both the frequency $\omega(k)$ and the group velocity of the modes are measured in the laboratory frame.
Panels (a,b): white hole configuration with a rightward flow $v_0>0$, as sketched in Fig.\ref{fig:sketch}(a). White hole parameters: $v_0/c_u=1.5$, $v_0/c_d=0.75$. Panels (c,d): black hole configuration with a leftward flow $v_0<0$, as sketched in Fig.\ref{fig:sketch}(b). Black hole parameters: $|v_0|/c_d=1.5$, $|v_0|/c_u=0.75$.
}
\label{fig:Bogo}
\end{center}
\end{figure}
\footnotetext{Note that this convention differs from the one that is usually adopted in relativity, where the modes are labelled 
according to their left- or right-moving character with respect to the light/sound cone, i.e. their propagation direction in the frame comoving with the condensate~\cite{macher2,macher1}. 
}

The white hole configuration that is the subject of the present paper is characterized by the chain inequality $c_u < v_0 < c_d$ relating the speed of sound in respectively the upstream and downstream asymptotic regions $c_{u,d} =\sqrt{\mu_{u,d}/m}$ and the (spatially uniform) flow speed $v_0$.
The dispersion of the Bogoliubov modes describing in the laboratory frame the propagation of weak disturbances on top of the flowing condensate has the usual form
\begin{equation}
\omega(k)=\Omega_{u,d}(k) 
+v_0 k=\pm \sqrt{\frac{k^2}{2m}\left(\frac{\hbar^2k^2}{2m}+2mc_{u,d}^2 \right)}+v_0 k
\eqname{bogo}
\end{equation}
in respectively the upstream and downstream regions and is illustrated in the upper (a,b) panels of Fig.\ref{fig:Bogo}. For future use, we have defined $\Omega_{u,d}(k)$ 
as the 
frequencies measured
in the upstream and down stream comoving frames at rest with the condensate. 

\begin{figure}[htbp]
\begin{center}
\includegraphics[width=0.7\columnwidth,angle=0,clip]{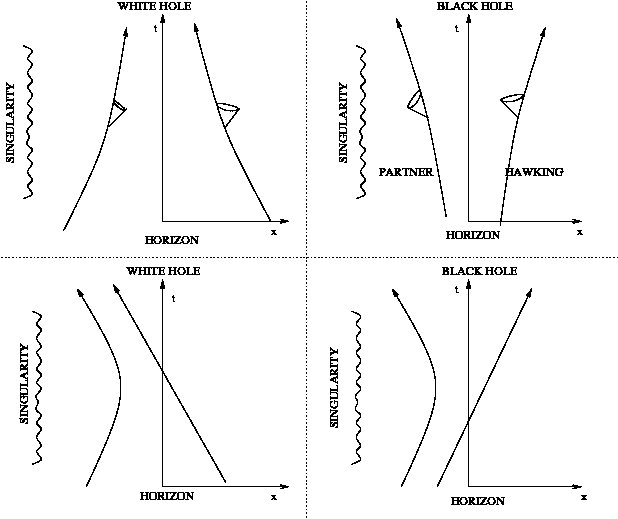}
\caption{
Sketch of the mode propagation around a horizon of the white (left panels) and black (right panels) hole type. Upper panels refer to the case of linear dispersion. Lower panels include the super-luminal contribution to the dispersion.
}
\label{fig:propag_sketch}
\end{center}
\end{figure}

The main feature of the white hole configuration consists of the fact that long wavelength phonons are forbidden from penetrating into the inner region of the white hole. As a consequence of dragging by the super-sonic flow, the group velocity turns out to be positive $v_0\pm c_u >0$ in the laboratory frame for both $k>0$ and $k<0$ phonons. This behaviour has to be compared to the black hole case shown in the lower panels of Fig.\ref{fig:Bogo}: in this case, the direction of the condensate flow is reversed and long wavelength phonons cannot escape from the inner, super-sonic region. It is however crucial to note that the above black and white hole behaviours are restricted to low-$k$ Bogoliubov excitations of hydrodynamic nature. Indeed, large-$k$ Bogoliubov excitations have a single particle character and propagate at a faster group velocity. As a result, they are able to propagate in both directions, that is to penetrate into a white hole and escape from a black hole. The transition from the hydrodynamic to the single-particle regime occurs at a characteristic wavevector given by the inverse of the healing length, $\xi_{u,d}=\hbar/m c_{u,d}$.

In terms of the gravitational analogy, the point separating the upstream region of super-sonic $c_u<v_0$ flow from the downstream one of sub-sonic $v_0<c_d$ flow then acts as the {\em horizon}.
The $k<0$ hydrodynamic phonons that propagate in the sub-sonic region in the leftward direction towards the horizon get slowed down as they approach the horizon and can never penetrate into the supersonic region. As a result, they pile up in the vicinity of the horizon. Analogously, the group velocity of $k<0$ hydrodynamic phonons in the super-sonic region tends to zero as the horizon is approached. As a result, they get also piled up on the horizon. The mechanism of this pile-up effect is graphically illustrated in the upper-left panel of Fig.\ref{fig:propag_sketch}.
As propagation on a stationary background conserves the phonon frequency $\omega$ in the laboratory frame, the pile-up effect is in both cases accompanied by an increase of the wavevector $k$ and a corresponding blue shift of the phonon frequency $\Omega(k)$ in the comoving frame. 
While this blue-shift effect continues for ever in the case of dispersionless (relativistic) fields giving rise to excitations at arbitrarily high $k$ values, in a Bose-Einstein condensate is restricted to wavevectors smaller than the inverse healing length by the peculiar super-luminal shape of the Bogoliubov dispersion shown in Fig.\ref{fig:Bogo}.

As the white hole configuration is related to the black hole one by a simple time reversal, the corresponding Bogoliubov dispersions shown in the upper (a,b) and lower (c,d) panels of Fig.\ref{fig:Bogo} are connected by a $k\rightarrow -k$ symmetry. From the figure, it is immediate to see that under this operation the sign of the group velocity is reversed and the in- and out-going character of the modes are exchanged.
Another, related difference involves the zero-mode defined by the $\omega(k_Z)=0$ condition. Such mode only exists in the super-sonic region and propagates opposite to the condensate flow: while in a black hole it propagates towards the horizon, in a white hole it propagates away from the horizon into the inner region. As a result, a perturbation of the white hole horizon from the Hartree condition \eq{Hartree} may result into the Landau-\u Cerenkov~\cite{Cerenkov} emission of phonons into this zero-energy mode of wavevector
\begin{equation}
k_Z=-\frac{2m}{\hbar}\sqrt{v_0^2-c_u^2}
\eqname{k0}
\end{equation}
comparable to the inverse healing length $\xi_{u,d}^{-1}$ and group velocity
\begin{equation}
v_Z=\frac{c_u^2-v_0^2}{v_0}
\eqname{v_k0}
\end{equation}
In the following of the paper, we shall see how this fact plays a crucial role in the physics of the white hole configuration.

\section{Dynamical stability of a white hole configuration: mean-field theory}
\label{sec:stab}

\subsection{Scattering of phonon wavepackets on the horizon}
\label{sec:wavepack}

\begin{figure}[htbp]
\begin{center}
\includegraphics[width=0.95\columnwidth,angle=0,clip]{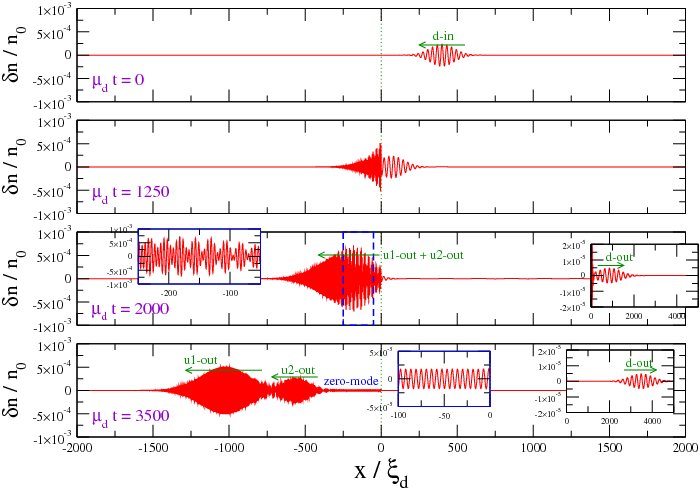}
\caption{
Scattering of a phonon wavepacket on a white hole horizon. The different panels show snapshots of the density modulation $\delta n(x)=n(x)-n_0$ at subsequent times. The insets provide a closer look at some most interesting features. White hole parameters as in Fig.\ref{fig:Bogo}(a,b) with $\sigma_{x}/\xi_d=0.5$. Wavepacket parameters $k_{wp}\xi_d=0.2$, $\sigma_{wp}/\xi_d=80$. The qualitative behavior remains unchanged if smoother horizons are considered with $\sigma_x/\xi_d\gg1$.
}
\label{fig:wavepacket}
\end{center}
\end{figure}

In the previous section we have seen that a condensate wavefunction in the plane wave form \eq{planewave} is at all times a solution of the GPE \eq{GPE}. Still, one has to assess the dynamical stability of this solution. A simple way to answer to this question consists of solving the Bogoliubov-de Gennes equations for the spectrum of linear perturbations on top of the flowing condensate and looking for complex frequency modes. Pioneering calculations in this direction were performed in~\cite{barcelo} for different black and white hole configurations and for different boundary conditions. A detailed investigation of black hole laser instabilities in combined black and white hole configurations was reported in~\cite{coutant,finazzi}. Application of this method to the white hole case is briefly discussed in Sec.\ref{sec:anal_stab}.

In the present Section, we shall follow a different method which consists in studying the wavefunction evolution under the full GPE \eq{GPE} for an initial condition slightly perturbed from the plane wave solution \eq{planewave}. This approach takes into account the nonlinear couplings that are neglected in the Bogoliubov theory and that are responsible for interesting back-reaction effects. In all calculations, the white hole is assumed to be already formed well before the experiment is performed so that the condensate background can be safely considered as a stationary one.

In particular, we consider the simple case where a phonon wavepacket modulation has been imprinted on top of the homogeneous solution \eq{planewave}. The wavepacket has a carrier at $k_{wp}$ and a Gaussian envelope of width $\ell_{wp}$ much larger than the wavelength, $k_{wp}\ell_{wp}\gg1$. Initially, the wavepacket is centered at a position $x_{wp}^o$ far from the horizon, $|x_{wp}^o|\gg \ell_{wp}$.  In wavevector space, the phonon wavepacket has a narrow Gaussian shape around $k_{wp}$ of width $\Delta k_{wp}=1/\ell_{wp}\ll k_{wp}$. The amplitude of the wavepacket is chosen to be weak enough to be in the linear regime; this condition has been checked by verifying that the amplitude of the reflected and transmitted wavepackets scales linearly with the amplitude of the incident wavepacket. A remarkable feature stemming from nonlinear effects beyond this approximation is discussed in Sec.\ref{sec:back}. 

Snapshots of the wavepacket evolution at subsequent times are given in Fig.\ref{fig:wavepacket} for the most illustrative case in which the wavepacket is initially at $x_{wp}^o>0$ outside the white hole and propagates towards the horizon on the $d^{\mathrm in}$ mode with a wavevector $k_{wp}<0$ well within the hydrodynamic region, $|k_{wp}|\ll \xi_d^{-1}$. Its frequency distribution is concentrated around the carrier frequency $\omega_{wp}\simeq (v_0-c_d)\,|k_{wp}|$. As the wavevector distribution is concentrated in the hydrodynamic region $k\xi_d\ll 1$ where the dispersion is almost linear, the wavepacket gets weakly distorted while propagating towards the horizon. Dynamical stability of the white hole configuration is confirmed by the fact that no exponentially growing perturbation of the horizon is triggered by the arrival of the incident wavepacket.

Within a strict hydrodynamic picture where the high momentum $u_{1,2}^{\rm out}$ modes were completely neglected, one would expect that no excitation can propagate across the horizon into the white hole as no leftward propagating modes are available inside the white hole at $\omega_{wp}$. As a consequence, the incident wavepacket either accumulates at the horizon or is reflected.
This picture is oversimplified as the combination of the blue shift at the white hole horizon and the super-luminal character of the Bogoliubov dispersion may lead to a significant transmission into $u_{1,2}^{\rm out}$ modes with large wavevectors in the vicinity of $\pm k_Z$. In~\cite{macher2}, it was predicted that the amplitude of these transmitted wavepackets is much larger than the one of the reflected one even for very smooth white hole horizons.

These expectations are confirmed in our numerical calculations shown in Fig.\ref{fig:wavepacket}. The reflected wavepacket is actually small, but clearly visible in the insets of the two latest time panels. As expected, it propagates away from the horizon at the speed $c_d+v_0$. In agreement with the condition that the carrier frequency $\omega_{wp}$  has to be conserved after scattering, the carrier wavevector of the reflected wave packet is smaller than the incident one by a factor $(v_0-c_d)/(v_0+c_d)$.

At relatively short times after impinging on the horizon (see e.g. the $\mu_d t=2000$ panel of Fig.\ref{fig:wavepacket}), there is a single transmitted wavepacket that propagates into the white hole with a group velocity close to $v_Z$. As one can see in the left inset, its shape is characterized by a carrier at a wavevector $k_Z$ and a slower modulation.
This latter can be interpreted as the result of interference between the $u_1^{\mathrm out}$ and $u_2^{\mathrm out}$ components of wavevector approximately $k_{u_1^{\rm out}}\simeq k_Z+\omega_{wp}/v_Z$ and $k_{u_2^{\rm out}}\simeq -k_Z+\omega_{wp}/v_Z$, see Fig.\ref{fig:Bogo}(a).
From the Bogoliubov expression for the density perturbation~\cite{castin}, one immediately recognizes oscillations at $\pm k_{u_1^{\rm out}}$ and $\pm k_{u_2^{\rm out}}$. The slow modulation corresponds to the beating of the $\pm k_{u_1^{\rm out}}$ and $\mp k_{u_2^{\rm out}}$ components, and has a wavevector $q_{beat}\approx 2\omega_{wp}/|v_Z|\approx 2(c_d-v_0)\,|k_{wp}|/|v_Z|$. The quantitative agreement of this Bogoliubov prediction for $q_{beat}$ with the period of the slow modulation extracted from the $\mu_d t=2000$ panel of Fig.\ref{fig:wavepacket} confirms our interpretation.

At later times (e.g. $\mu_d t=3500$), the curvature of the Bogoliubov dispersion makes the $u_1^{\mathrm out}$ and $u_2^{\mathrm out}$ wavepackets to eventually separate in space. As a result, the slow modulation due to their interference disappears and one is left with two Gaussian wavepackets. The presence of the transmitted wavepacket on the negative norm $u_{2}^{\mathrm out}$ out-going mode can be interpreted as the result of a stimulated Hawking emission~\cite{PRA_gradino}. This classical counterpart of Hawking radiation was recently investigated in experiments using gravity waves in water tanks~\cite{Weinfurtner,Rousseaux}.

Had we considered an incident wave packet with a smaller wave vector $k_{wp}$, the time needed for the two transmitted wave packets to separate would be longer as the group velocities associated with the two roots  $u_1^{\rm out}$ and $u_2^{\rm out}$ would be closer. In the $|k_{wp}|\to 0$ limit, the two wave vectors $k_{u_1}^{\rm out}$ and $k_{u_2}^{\rm out}$ tend to $\pm k_Z$ and the wavepackets never separate.

\begin{figure}[htbp]
\begin{center}
\includegraphics[width=0.95\columnwidth,angle=0,clip]{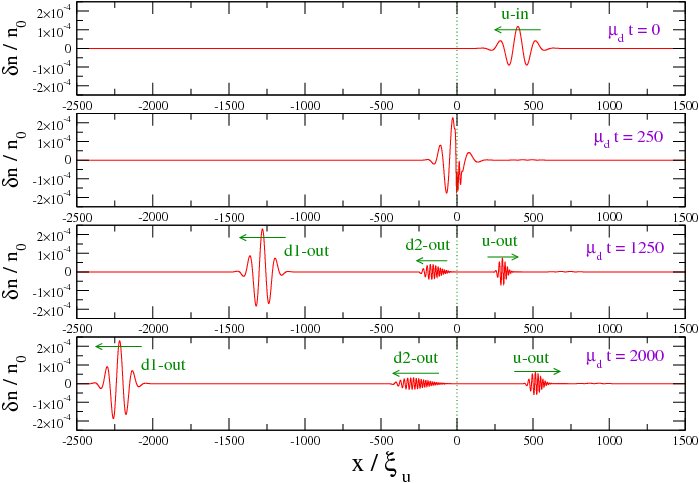}
\caption{
Scattering of a phonon wavepacket on a black hole horizon. The different panels show snapshots of the density modulation $\delta n(x)=n(x)-n_0$ at subsequent times. Black hole parameters as in Fig.\ref{fig:Bogo}(c,d) with $\sigma_x/\xi_u=0.5$. Wavepacket parameters $k_{wp}\xi_u=0.05$, $\sigma_{wp}/\xi_u=80$.
}
\label{fig:wpBH}
\end{center}
\end{figure}

For the sake of completeness, it is interesting to compare these results to the case of a black hole configuration. As before, we consider a Gaussian wavepacket centered at $x_{wp}^o>0$ outside the black hole which propagates towards the horizon on the $u^{\mathrm in}$ mode with a wavevector $k_{wp}<0$ well within the hydrodynamic region, $|k_{wp}|\ll \xi_u^{-1}$.
As one can see in the snapshots shown in Fig.\ref{fig:wpBH}, the $u^{\mathrm in}$ incident wave packet scatters on the horizon and splits into a reflected wavepacket on the $u^{\mathrm out}$ mode and a pair of transmitted ones on the modes labelled as $d_1^{\mathrm out}$ and $d_2^{\mathrm out}$ in Fig.\ref{fig:Bogo}(c,d): while the stronger and faster wavepacket corresponds to the transmitted
positive-norm $d_1^{\mathrm out}$ mode, the weaker one corresponds to the negative-norm $d_2^{\mathrm out}$ mode and results from the same mixing of positive and negative modes that underlies Hawking radiation~\cite{PRA_gradino}. In contrast to the white hole case, all out-going wave packets have significantly different group velocities and separate in space almost immediately; furthermore, the carrier wavevectors of all wavepackets tend to zero in the $|k_{wp}|\to 0$ limit. The reason of these remarkable differences is easily understood in terms of the dispersion curves shown in Fig.\ref{fig:Bogo}.

\subsection{Back-reaction of the wave packet on the horizon}
\label{sec:back}

\begin{figure}[htbp]
\begin{center}
\includegraphics[width=0.75\columnwidth,angle=0,clip]{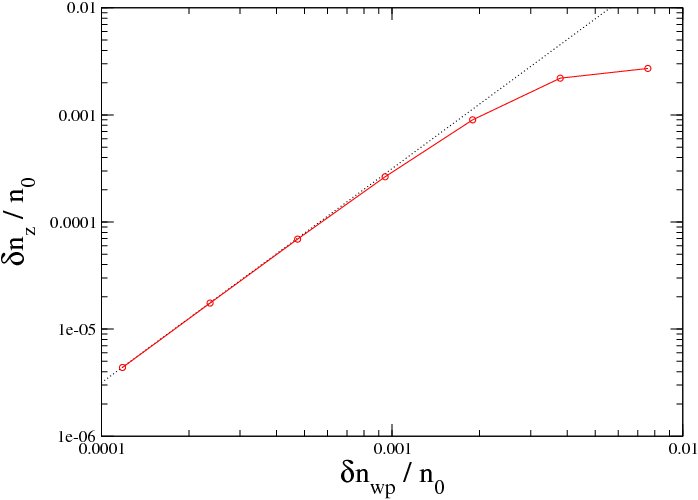}
\caption{
Log-log plot of the amplitude of the zero-mode modulation as a function of the peak incident wavepacket amplitude. The zero-mode modulation amplitude is extracted from late-time snapshots of simulations of the same configuration as considered in Fig.\ref{fig:wavepacket}. The black dotted line is a fit with a square law $\delta n_{z}= C\,\delta n_{wp}^2$.
}
\label{fig:check_backreaction}
\end{center}
\end{figure}

The previous discussion of phonon scattering on the white hole horizon has highlighted a quite similar physics as compared to the black hole case: an incident wave packet splits in several reflected and transmitted wavepackets whose carrier wavevector is fixed by energy conservation arguments. As we are considering weak incident wavepackets, the amplitude of the reflected and transmitted scales linearly with the incident one.
However, in addition to these expected wave packets, a striking unexpected feature is visible in the latest time snapshot of Fig.\ref{fig:wavepacket} right inside the white hole horizon: a stationary density modulation with a spatially flat envelope and a fast oscillating carrier. At late times, the amplitude of this density modulation saturates to a finite value. A similar effect was very recently reported for a classical white hole analog using gravity waves on the surface of a water tank~\cite{Weinfurtner,Rousseaux}.

Its physical origin can be understood in terms of the zero-mode in the Bogoliubov dispersion: the vanishing $\omega$-frequency
of this mode explains the stationarity of the pattern and the observed wavelength of the modulation coincides with the value $k_Z$ of the zero-mode wavevector in the upstream region. Still, the microscopic rectification process that is responsible for the excitation of a zero-frequency mode from a finite frequency wavepacket requires further investigation as it is not allowed within a linearized Bogoliubov theory on a stationary background.

To assess its origin in terms of some nonlinear effect, we have repeated the same numerical simulation of the GPE with different amplitudes of the incident wavepacket. The results are summarized in Fig.\ref{fig:check_backreaction}: the late-time amplitude of the zero-mode modulation scales indeed proportionally to the {\em square} of the incident wavepacket amplitude. A simplest physical interpretation can be put forward as follows: the incident wavepacket slightly distorts the wavefunction in the horizon region breaking the Hartree condition \eq{Hartree}. The resulting perturbation of the Hartree potential is then responsible for the continuous emission of phonons via the Landau-\u Cerenkov emission discussed in~\cite{Cerenkov,hakim} for a uniform super-sonic flow hitting a defect. 

This interpretation is confirmed by the numerical observation that no such feature was ever observed for black hole configurations: in this latter case, the super-sonic region is in fact located downstream of the defect so that the zero-mode propagates towards the horizon. This fact is clearly visible in Fig.\ref{fig:Bogo}(c). As a result, the zero-mode can not be excited by a perturbation localized in the horizon region of a black hole.

\vspace{0.5cm}

To summarize, our numerical simulations of the Gross-Pitaevskii equation show that the white hole configuration is dynamically stable at mean-field level even beyond the linearized regime considered so far in the literature. Even when it is excited by a sizable incident wavepacket, the condensate does not develop any exponentially growing perturbation: the amplitude of the scattered wavepackets is roughly proportional to the incident amplitude and the amplitude of the zero-mode perturbation tends at late times to a stationary value proportional to the square of the incident wavepacket amplitude.

\section{The effect of quantum fluctuations}
\label{sec:quantumfluct}

After having assessed in the previous Section the dynamical stability of a white hole configuration at the level of mean-field theory, we now proceed to investigate the consequences of quantum fluctuations, and in particular the radiation that is continuously emitted by the white hole horizon once this has been formed. Our investigation will make use of the same truncated-Wigner numerical technique used in~\cite{NJP_BH}. We refer to this paper for a more detailed discussion about technical details.

At the initial time $t=0$, the condensate is assumed to be spatially homogeneous at density $n_0$ and to be flowing at a constant speed $v_0$; both the interaction constant and the external potential are initially flat and equal to $g_d$ and $V_d$, respectively. The white hole horizon is formed at the time $t_0$ by ramping $g$ and $V$ in the upstream $x<0$ region to their final values $g_u$ and $V_u$. This ramp takes place within a time $\sigma_t$. Eventually, $g$ and $V$ tend to a arctan-shaped spatial profile of characteristic thickness $\sigma_x$.

Quantum and thermal fluctuations on top of the initial homogeneous condensate can be treated by means of the standard Bogoliubov theory of dilute Bose gas. Within the Wigner method, this translates~\cite{WMC} into a stochastic initial condition with a suitable Gaussian noise added on top of the plane-wave initial wavefunction \eq{planewave} of mean-field theory. The variance of the Gaussian noise is determined by the initial temperature $T_0$. Exception made for Fig.\ref{fig:cuts2}(b), all numerical data shown in the figures refer to the case of a vanishing initial temperature $T_0=0$.
The time evolution of the stochastic wavefunction starting from its initial state follows the deterministic GPE evolution \eq{GPE} including the chosen form of the nonlinear interaction constant $g(x,t)$ and of the external potential $V(x,t)$. Expectation values of symmetrically ordered observables at any later time are obtained by taking the corresponding stochastic average over the ensemble of evolved wavefunctions.

\subsection{Signatures of white hole emission in the correlation pattern of density fluctuations}
\begin{figure}[htbp]
\begin{center}
\includegraphics[width=0.8\columnwidth,angle=0,clip]{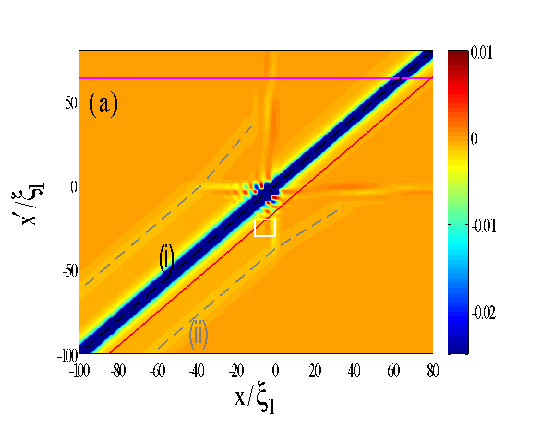}
\includegraphics[width=0.8\columnwidth,angle=0,clip]{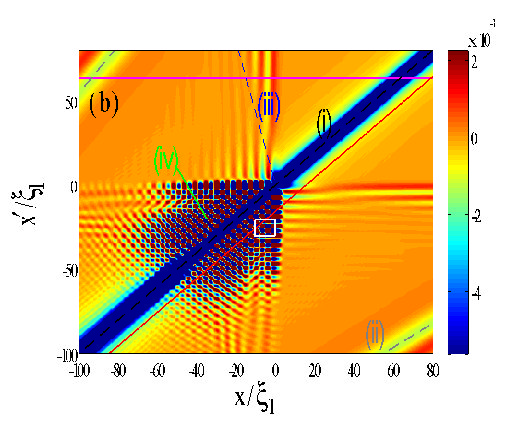}
\caption{Color plots of the rescaled density correlation function $(n_0\xi_d)\times[G^{(2)}(x,x')-1]$ at two successive times $\mu_d t=90$ (a) and $160$ (b) after the switch-on of the white hole horizon.
The $(i)$, $(ii)$, $(iii)$, $(iv)$ labels identify the main features discussed in the text.
The solid magenta and red lines indicate the directions along which the cuts shown in Fig.\ref{fig:cuts1}(a,b) are taken. The white rectangle indicate the region where the checkerboard amplitude shown in Fig.\ref{fig:cuts2} is measured. The blue dashed line in panel (b) indicates the analytical position of the axis of feature (iii).
The horizon is formed within a time $\mu_d \,\sigma_t=10$ around $\mu_d t_0=50$ and has a spatial thickness $\sigma_x/\xi_d=0.5$. White hole parameters: $v_0/c_u=1.5$, $v_0/c_d=0.75$. The initial temperature is $T_0=0$.
}
\label{fig:g2}
\end{center}
\end{figure}

\begin{figure}[htbp]
\begin{center}
\includegraphics[width=0.80\columnwidth,angle=0,clip]{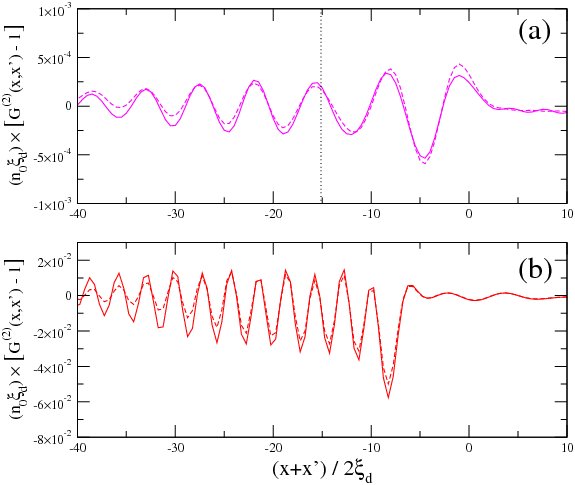}
\caption{Cuts of the rescaled density correlation function at times $\mu_1 t=160$ (dotted lines), $180$ (solid lines) taken along respectively the (a) magenta ($x'=63.5\xi_d$) and (b) red ($x-x'=15.5\,\xi_d$) lines of Fig.\ref{fig:g2}. The dashed line in (a) indicates the prediction \eq{center_inout} for the central position of the $d-u_{1,2}$ correlation discussed in Sec.\ref{sec:corr_inout}.
}
\label{fig:cuts1}
\end{center}
\end{figure}

\begin{figure}[htbp]
\begin{center}
\includegraphics[width=0.80\columnwidth,angle=0,clip]{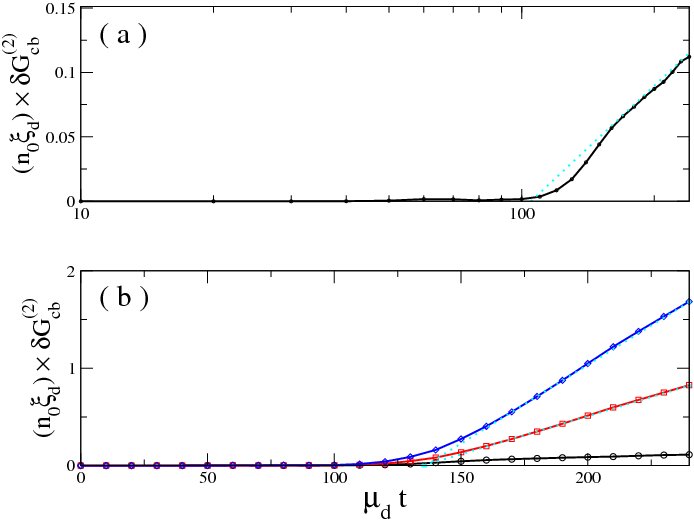}
\caption{Time evolution of the peak-to-peak amplitude of the checkerboard pattern measured within the region indicated by the white rectangle in Fig.\ref{fig:g2}. The upper (a) panel shows a semilog plot for a zero initial temperature $T_0=0$. The lower (b) panel shows a linear plot for different temperatures $k_B T_0/\mu_d=0$ (black), $0.1$ (red) and $0.2$ (blue). The cyan straight lines are guides to the eye which emphasize that the growth is approximatively logarithmic (linear) in the zero (finite) temperature cases.
}
\label{fig:cuts2}
\end{center}
\end{figure}

In~\cite{NJP_BH} it was shown that
a good deal of information of the physical properties of Hawking radiation 
can be extracted from the correlation pattern of density fluctuations at equal times, defined as
\begin{eqnarray}
G^{(2)}(x,x')&=&\frac{\langle \Psihd(x)\,\Psihd(x')\,\Psih(x')\,\Psih(x)\rangle }{\langle \Psihd(x)\,\Psih(x) \rangle\, \langle \Psihd(x')\,\Psih(x') \rangle}= \nonumber \\
&=& \frac{\langle n(x) n(x') \rangle }{\langle n(x) \rangle \, \langle n(x') \rangle} - \frac{1}{\langle n(x) \rangle} \, \delta(x-x'). 
\end{eqnarray}
In this section, we follow the same line for the case of a white hole, trying to isolate the peculiar features that characterize the white hole counterpart of the Hawking radiation from black holes.
Looking at the Bogoliubov dispersions shown in Fig.\ref{fig:Bogo}(a,b), we expect from energy conservation arguments that correlated pairs of quanta can be emitted from the horizon either into $u_1^{\rm out}$ and $u_2^{\rm out}$ modes or into the $d^{\rm out}$ and $u_2^{\rm out}$ modes: the negative energy of the $u_2^{\rm out}$ partner can in fact compensate the positive energy of the $u_1^{\rm out}$ or $d^{\rm out}$ ones, so to give a zero total energy of the pair.
As it happened in black holes, these processes correspond to clearly distinct features in the correlation pattern. An analytical understanding of this {\em white hole radiation} will be provided in Sec.\ref{sec:analytic}. Color plots of a suitably normalized density correlation $G^{(2)}(x,x')$ are shown in Fig.\ref{fig:g2}.  For numerical convenience, all density correlation plots have been smoothened out with a Gaussian spatial filter of size $\ell_{av}/\xi_d=1$. 
The main features that are visible in the plots can be classified as follows:

\begin{itemize}

\item[$(i)$] The strong and negative correlation strip on the $x=x'$ main diagonal results from the usual many-body antibunching due to repulsive interactions and has no relation with the presence of an horizon.

\item[$(ii)$] A system of stripes parallel to the main diagonal appears in the $x<0$ region inside the white hole as soon as the horizon is formed. As time goes on, the fringes move away from the main diagonal at an approximately constant speed and eventually disappear from the region of sight. We have checked that their presence and shape does not depend on the presence of the horizon, but rather on the speed of the time-modulation: the slower the ramp (i.e. the longer $\sigma_t$), the weaker the fringe amplitude. On the basis of~\cite{DCE}, these fringes can be interpreted as a consequence of the dynamical Casimir emission of pairs of counterpropagating phonons at all points where the interaction constant is modulated in time, $g=g_d\rightarrow g_u$. It is interesting to note that in the white hole case the dynamical Casimir phonons are able to travel back to the horizon. As a direct consequence of this fact, feature (ii) extends also into the $x>0$, $x'<0$ and $x<0$, $x'>0$ quadrants. As the scattering of dynamical Casimir phonons incident off the horizon may interfere with the white hole emission, it is then important to ensure that the intensity of the dynamical Casimir emission is efficiently suppressed by a suitably long formation time $\sigma_t$.

\item[$(iii)$] Spatially oscillating correlations appear between points right outside and right inside the horizon. The correlation pattern has a sinusoidal dependence as a function of the position of the point located inside the horizon, while the dependence on the outside point is much smoother.
The wavevector of the fringes is approximately equal to the zero-mode wavevector $k_Z$ introduced in the previous section.
As time goes on, these fringes extend into a larger area further away from the horizon, but their amplitude at each given point eventually tends to a finite value, constant in time. This last fact is better visible in the cuts taken along the magenta line that are shown in Fig.\ref{fig:cuts1}(a).
Remarkably, the envelope of the fringe system shows a slow modulation, with a minimum approximately located along the blue dashed line of Fig.\ref{fig:g2}(b), $x/v_Z=x'/(v_0+c_d)$, which suggests that this feature is related to the emission of pairs of quanta by the horizon into the $u_{1,2}$ and $d$ modes. An analytical discussion of this feature will be provided in Sec.\ref{sec:corr_inout}.

\item[$(iv)$] A checkerboard pattern appears in the correlations between pair of points located inside the white hole. As time goes on, the checkerboard pattern extends further away from the horizon. The spatial wavevector of the checkerboard is approximately equal to $k_Z$ in both $x,x'$ directions, which suggests that it is due to the emission of pairs of quanta by the horizon into the $u_{1,2}$ modes.
In contrast to feature $(iii)$ and to what was always observed for black holes in~\cite{NJP_BH}, its amplitude at any given point does not tend to a stationary value, but keeps on steadily increasing.
This remarkable fact is visible in the cuts of the correlation pattern shown in Fig.\ref{fig:cuts1}(b) and, even more clearly, in the plot of the checkerboard amplitude as a function of time shown in Fig.\ref{fig:cuts2}(a,b). For a vanishing initial temperature $T_0=0$, the growth appears to follow a logarithmic law. For a finite initial temperature $T_0>0$, the growth appears to be instead linear with a slope proportional to $T_0$.
An analytical interpretation of all these features will be discussed in Sec.\ref{sec:corr_inin}.

\end{itemize}

\subsection{Comparison with Hawking radiation from black holes}

\begin{figure}[ptb]
\begin{center}
\includegraphics[width=0.44\columnwidth,angle=0,clip]{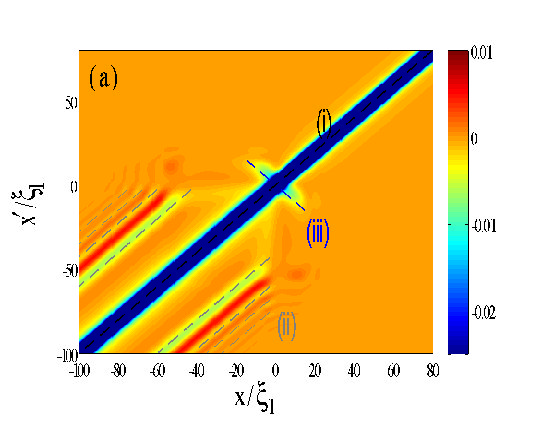}\hspace{0.1\columnwidth}
\includegraphics[width=0.44\columnwidth,angle=0,clip]{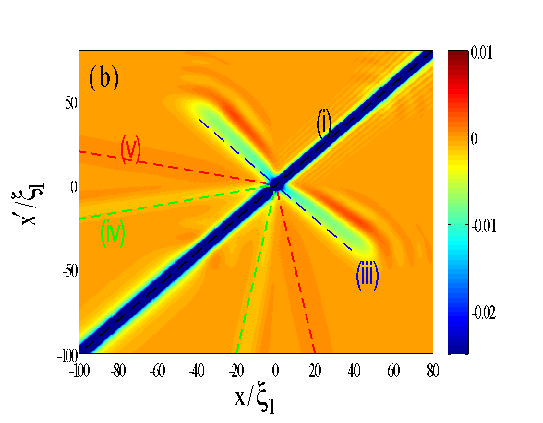}
\caption{Color plots of the rescaled density correlation function $(n_0\xi_d)\times[G^{(2)}(x,x')-1]$ at two successive times $\mu_u t=40$ (a) and $160$ (b) after the switch-on of a black hole horizon.
The $(i)$, $(ii)$, $(iii)$, $(iv)$ labels identify the main features discussed in the text.
The horizon is formed within a time $\mu_u \,\sigma_t=0.5$ around $\mu_u\,t_0=2.5$ and has a spatial thickness $\sigma_x/\xi_u=0.5$. The calculation has been performed using the truncated-Wigner method. Same system parameters as in Fig.\ref{fig:Bogo}(c,d). The initial temperature is $T_0=0$.}
\label{fig:g2BH}
\end{center}
\end{figure}

For the sake of completeness it can be useful to briefly review the main properties of the Hawking emission from a black hole~\cite{NJP_BH,PRA_gradino}. This should facilitate the reader in appreciating the analogies and differences with the white hole radiation. As before, this study is based on the correlation pattern of density fluctuations, some examples of which are shown in Fig. \ref{fig:g2BH}. Several features can be identified: 
\begin{itemize}

\item[$(i)$] As expected, the strong and negative correlation strip on the $x=x'$ main diagonal is almost identical in the white hole and black hole configurations.

\item[$(ii)$] The moving pattern of fringes due to the dynamical Casimir emission of phonons by the time-modulated interaction constant is also very similar in the black and white hole configuration. The main difference is purely geometrical: as a result of the different relative position of the super- and sub-sonic regions, the low-$k$ dynamical Casimir phonons are not able to travel back to the horizon and are fully dragged into the black hole.

\item[$(iii)$] A symmetric pair of negative correlation tongues extend from the horizon point for points $x$ and $x'$ located on opposite sides from the horizon. Their axis is located along the $x/(v_0+c_d)=x'/(v_0+c_u)$ straight line.
While their length linearly grows with time since the horizon formation time, their maximum height remains almost constant in time. Their physical origin lies in the stationary emission of correlated pairs of quanta by the horizon into the modes labelled as $u^{\rm out}$ and $d_2^{\rm out}$ in Fig.\ref{fig:Bogo}(c,d): while the $d_2$ partner falls into the black hole, the $u$ partner flies away giving rise to the standard Hawking emission.

\item[$(iv)$] Another pair of positive correlation tongues extends from the horizon point for both $x,x'<0$ inside the horizon. Their axis is located along the $x/(v_0+c_d)=x'/(v_0-c_d)$ line, which suggests an interpretation in terms of the stationary emission of correlated pairs of $d_1^{\rm out}$ and $d_2^{\rm out}$ quanta by the horizon.

\item[$(v)$] Another pair of weaker tongues appears in the $x<0$, $x'>0$ region along the $x/(v_0-c_d)=x'/(v_0+c_u)$ line, which suggests an interpretation in terms of correlations between the $d_1^{\rm out}$ and $u^{\rm out}$ modes.

\end{itemize}

A general fact of all (iii)-(v) features is that none of them shows any significant modulation: this is due to the fact that the wavevector of all out-going phonon modes tends to zero in a black hole configuration for low frequencies. This is to be contrasted to the white hole case where the wavevector of $u_{1,2}^{\rm out}$ modes tends to $\pm k_Z$ which is of the order of the inverse healing length $1/\xi_u$ in the supersonic region.
\vspace{0.5cm}

To summarize, in this Section we have described the properties of the quantum emission of phonons from the white hole horizon. As in the black hole case, the peculiar quantum correlations between the consituents of each pairs result into long distance correlations of the density fluctuations. Several novel features have been highlighted: as the quantum emission into the white hole consists of short wavelength excitations comparable to the healing length, the correlation pattern shows a fast sinusoidal modulation. Furthermore, the amplitude of the checkerboard pattern for pairs of points located inside the white hole keeps on growing in time. This is in stark contrast with the stationary character of Hawking radiation from black holes. The consequence of this steady growth of density fluctuations for the stability of the white hole configuration requires a detailed study of the back-action effect of quantum fluctuations onto the flow and will be the subject of further work.

\section{Analytical theory}
\label{sec:analytic}

In this Section we sketch an analytical framework offering a physical interpretation of the numerical observations discussed in the previous Sections. More details will be given in future publications, in particular in Carlos Mayoral's PhD thesis~\cite{Carlos_phd}.
Our approach is based on the step-like horizon model first introduced at the level of mean-field theory in~\cite{barcelo} and then developed into a full theory of quantum fluctuations and Hawking radiation in~\cite{PRA_gradino}. Here, we will show that this theoretical model is able to explain most of the numerically observed physics and to provide quantitatively accurate predictions for most relevant observables.

\subsection{Dynamical stability of white holes}
\label{sec:anal_stab}

In this first subsection we address the issue of the dynamical stability of the white hole configuration. The numerical observations reported in Sec.\ref{sec:stab} indicate that the chosen configuration is dynamically stable. Here we confirm this conclusion by analytical means. Our analytical approach is inspired by~\cite{barcelo} and gets to similar conclusions.
For analytical simplicity, we restrict our attention to the same configuration used in the numerical calculations, that is a BEC in the plane wave state \eq{planewave} with a spatially constant density $n_0$ and flow speed $v_0$; the step-like white hole horizon is created by means of a sudden jump at $x=0$ of both the nonlinear interaction constant $g(x)$ and the external potential $V(x)$.

Dynamical stability of this configuration against perturbations of the horizon region is assessed by looking for eigenmodes of the Bogoliubov-de Gennes equations~\cite{BECbook,castin} with complex frequency: the system is unstable if at least an eigenmode $\alpha$ exists with a frequency $\tilde{\omega}_\alpha$ such that $\textrm{Im}[\tilde{\omega}_\alpha]>0$. As we are interested in instabilities that start in the horizon region, boundary conditions have to imposed that the mode wavefunction of the perturbation decays to zero at infinity on both sides of the horizon. At the horizon, the mode wavefunction has to be continuous along with its first spatial derivative.

More precisely, we are looking for eigenmodes of the Bogoliubov-de Gennes equations:
\begin{eqnarray}
\hbar\tilde{\omega}u(x)&=&-\frac{\hbar^2}{2m}\frac{\partial^2}{\partial x^2}u(x)+V(x)u(x) + 2 g(x) |\phi_0|^2 u(x) + \phi_0^2(x) v(x) - \mu u(x) \eqname{BdG1}, \\
\hbar\tilde{\omega}v(x)&=&\frac{\hbar^2}{2m}\frac{\partial^2}{\partial x^2}v(x)-V(x)v(x) - 2 g(x) |\phi_0|^2 v(x) - \phi_0^2(x) u(x)+ \mu v(x) \eqname{BdG2}.
\end{eqnarray}
Within the spatially homogeneous upstream $x<0$ region, the eigenmodes can be expanded into plane waves as follows
\begin{eqnarray}
u(x)=\sum_i \alpha_i \tilde{u}^{(i)}_{u}\,e^{i\tilde{k}_u^{(i)} x} \eqname{uc} \\
v(x)=\sum_i \alpha_i \tilde{v}^{(i)}_{u}\,e^{i\tilde{k}_u^{(i)} x}, \eqname{vc}
\end{eqnarray}
where the sum over $i$ runs over the Bogoliubov modes at frequency $\tilde{\omega}$, whose complex wavevector $\tilde{k}_{u}^{(i)}$ satisfies the (complex) dispersion relation
\begin{equation}
\hbar^2 (\tilde{\omega}-\tilde{k}_u v_0)^2=\frac{\hbar^2 \tilde{k}_u^2}{2m}\left(\frac{\hbar^2 \tilde{k}_u^2}{2m}+ 2 g_u |\phi_0|^2 \right)
\eqname{bogo_disp_complex}
\end{equation}
describing Bogoliubov modes in a moving fluid at $v_0$. As usual, the amplitudes $\tilde{u}_u$ and $\tilde{v}_u$ are determined by inserting the plane wave ansatz \eqs{uc}{vc} into the Bogoliubov-de Gennes equations \eqs{BdG1}{BdG2}. Differently from the standard cases~\cite{BECbook,castin}, the complex value of $\tilde{\omega}$ makes the $\tilde{u},\tilde{v}$ coefficients to be complex as well.
\begin{figure}[htbp]
\begin{center}
\includegraphics[width=0.75\columnwidth,angle=0,clip]{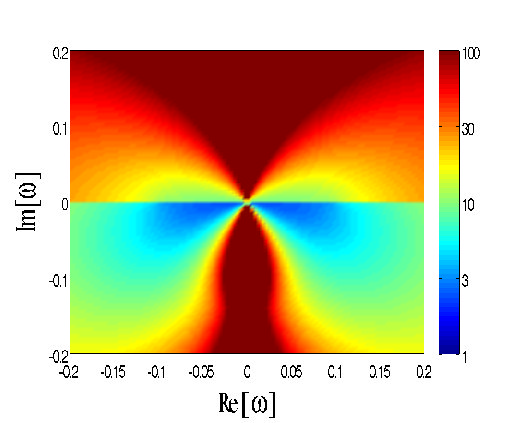}
\caption{Log-scale color plot of the determinant whose zeros define the asymptotically bound modes of a white hole configuration with a step-like horizon. X- and y-axis are the real and imaginary parts of the complex frequency $\tilde{\omega}$. As the determinant has no zeros in complex $\tilde{\omega}$ plane, no asymptotically bound modes exist. This proves {\em a fortiori} that the white hole is dynamically stable. Same white hole parameters as in the previous figures.
}
\label{fig:analytic_stab_WH}
\end{center}
\end{figure}

The boundary condition that the mode wavefunction has to decay to zero for $x\rightarrow -\infty$ restricts the sum to those Bogoliubov modes whose wavevector $\textrm{Im}[\tilde{k}^{(i)}_u]<0$: out of the four solutions of \eq{bogo_disp_complex}, only two can then be retained~\footnote{It can be shown that the number of such solutions is always two. More details on this issue as well as on the relation between this boundary condition and the in-going vs. out-going character of the modes will be given in future publications.}. Analogous conditions hold for the downstream, $x>0$ region.  The matching condition at the horizon imposes that both $u(x)$ and $v(x)$ have to be continuous, as well as their first spatial derivatives.

The boundary conditions at infinity plus the matching conditions at the horizon reduce our problem to the detemination of four amplitude coefficients (two modes per $u,d$ side) that have to satisfy four matching conditions. Non-trivial solutions (the so-called asympotically bound modes (ABM)~\cite{coutant,finazzi}) exist only for special values of the complex frequency $\tilde{\omega}$ that make the determinant of the corresponding linear system to vanish. As one can see in Fig.\ref{fig:analytic_stab_WH}, this never happens for the white hole system under investigation here. As dynamical instabilities are signaled by the existence of ABMs with $\textrm{Im}[\tilde{\omega}]>0$, the complete absence of ABMs confirms the numerical observation of Sec.\ref{sec:stab} that white holes are dynamically stable objects. Of course, the linearized Bogoliubov theory that is used in this Section is not able to capture the stationary Bogoliubov-\u Cerenkov density modulation pattern discussed in Sec.\ref{sec:back}, that appears as a consequence of nonlinear back-action effects.

In a forthcoming paper this theory will be extended to more complex configuration involving a pair of black and white hole horizons: in agreement with~\cite{coutant,finazzi}, the {\em black hole laser} phenomenon shows up as a series of discrete eigenmodes with a positive imaginary part of the frequency. Each of them corresponds to a zero of the determinant at the corresponding complex frequency $\tilde{\omega}$. In a figure analogous to Fig.\ref{fig:analytic_stab_WH}, such modes would appear as narrow blue spots corresponding to sharp minima of the logarithm.

\subsection{The $S$ matrix and the white hole radiation}

In the previous subsection we have shown that no complex frequency eigenmodes appear in a white hole configuration. All the physics of quantum fluctuations can then be studied in terms of the input-output formalism of quantum optics~\cite{qo,cc}: the central object of this approach is the ${S}(\omega)$ matrix connecting the amplitude operators for the out-going Bogoliubov modes to those of the in-going modes at the same real frequency $\omega$.  This is a direct extension to the white hole configuration of the Bogoliubov theory of acoustic black holes first introduced in~\cite{leonhardt_bogo} and fully developed in~\cite{macher2,macher1,PRA_gradino}.

A sketch of the in-going and out-going modes for the white hole configuration was shown in Fig.\ref{fig:Bogo}(a,b): in the low-energy sector below the cut-off frequency $\omega_{\rm max}$ (defined as the maximum of the dispersion of the $u_2$ mode), there are three in-going modes, labelled $d^{\mathrm in}$, $u_1^{\mathrm in}$ and $u_2^{\mathrm in}$, and three out-going modes, labelled $d^{\mathrm out}$, $u_1^{\mathrm out}$ and $u_2^{\mathrm out}$.
As it was discussed in previous works, the $d$ and $u_1$ out-going (in-going) modes have a positive Bogoliubov norm and enter in the input-output formalism as annihilation operators $\bh_{d,u_1}(\omega)$ ($\ah_{d,u_1}(\omega)$). On the other hand, the negative norm $u_2$ modes enter the formalism as creation operators $\bhd_{u_2}(\omega)$ ($\ahd_{u_2}(\omega)$). The input-output relations have the form:
\begin{equation}
\left(
\begin{array}{c}
\bh_d(\omega) \\ \bh_{u_1}(\omega) \\ \bhd_{u_2}(\omega)
\end{array}
\right)=
S(\omega)\,
\left(\begin{array}{c}
\ah_d(\omega) \\ \ah_{u_1}(\omega) \\ \ahd_{u_2}(\omega)
\end{array}
\right).
\eqname{S}
\end{equation}
The white hole radiation consists of a finite occupation of the out-going modes even for a vacuum state of the in-going ones. In this framework, it appears as a direct consequence of the mixing of creation and destruction operators by the $S$-matrix \eq{S}. As usual in quantum optics~\cite{qo,cc}, such processes lead to a squeezed vacuum state as the output state of the modes after having scattered on the horizon.
Exception made for some remarks at the end of Sec.\ref{sec:corr_inin}, our investigation is focussed on the case of a $T_0=0$ initial state where the in-going modes are assumed to be in their vacuum state defined by $\ah_{d,u_1,u_2}\,|\textrm{vac}\rangle=0$.  
The population of the out-going $d$ and $u_1$ modes is proportional to $|S_{du_2}|^2$ and $|S_{u_1u_2}|^2$, respectively. Thanks to the unitarity of the $S$ matrix, the population of the negative frequency partners is proportional to $|S_{u_2u_2}|^2=|S_{du_2}|^2 + |S_{u_1u_2}|^2$, which confirms that phonons are created in pairs which carry no net energy.

A simple way to estimate the matrix elements of $S(\omega)$ is to look at the time evolution of incident wavepackets as done in Sec.\ref{sec:stab}, and to extract the reflection and transmission amplitudes from the intensity and phase of the out-going wavepackets. Alternatively, one can obtain the Bogoliubov modes at a given $\omega$ by solving the corresponding Bogoliubov-de~Gennes differential equations: a numerical solution of this problem was performed in~\cite{macher2} for smooth configurations of the flow velocity and the speed of sound. Closed analytical formulas for the transmission and reflection amplitudes of a black hole in the opposite limit of a step-like horizon were obtained in~\cite{PRA_gradino} and then in~\cite{sandrocarlos}: imposing appropriate boundary conditions at the horizon $x=0$, one is able to match the Bogoliubov plane-wave solutions within each homogeneous region for $x\gtrless 0$ and then build a global solutions of the Bogoliubov problem.

Here this last strategy is applied to a white hole configuration to obtain closed expressions for the matrix elements of $S(\omega)$ in the low-frequency limit within the step-like horizon approximation. In particular, we shall focus our attention on the matrix elements for the negative frequency $u_2$ in-going mode that are involved in the white hole radiation:
\begin{eqnarray}
|S_{u_1u_2}|^2&\simeq& |S_{u_2u_2}|^2\simeq \frac{(v_0-c_d)(v_0^2-c_u^2)^{3/2}(c_d+c_u)}{2c_u(c_d+v_0)(c_u-c_d)}\,\frac{1}{\omega} \\
|S_{du_2}|^2&\simeq&\frac{c_d}{c_u}\frac{(v_0-c_u)^2}{(v_0+c_d)^2}.
\end{eqnarray}
This behavior of the low-energy $S$ matrix elements is common to all in-going modes: all the matrix elements for out-going $u_1$ and $u_2$ modes diverge as $1/\sqrt{\omega}$, while the ones for out-going $d$ mode tends to a constant value. As a consequence, the spectral distribution of the quanta that are emitted into the $u_{1,2}^{\rm out}$ modes into the white hole has the usual $1/\omega$ thermal form of Hawking emission, while the radiation leaving the horizon has a flat spectrum. This remarkable fact was first noticed in~\cite{macher2} as a result of a full Bogoliubov calculation with a smooth horizon, and constitutes a crucial difference from the black hole case where the Hawking emission outside the horizon has a thermal $1/\omega$ spectrum.

For higher frequencies (but still below $\omega_{\rm max}$), the $S_{u_1u_2}$ matrix element quickly decreases to small values, while the $S_{u_2u_2}$ one keeps a sizable value as it ends up describing simple reflection processes. For a smooth horizon of thickness $\sigma_x\gg \xi$, these coefficients are expected to tend to respectively $0$ and $1$ with an exponential law of the canonical Hawking form
\begin{eqnarray}
|S^{(H)}_{u_2u_2}| &=& [ 1 - \exp(-\omega/\omega_H) ]^{-1/2}\\
|S^{(H)}_{u_1u_2}| &=& \exp(-\omega/2\omega_H)\,|S_{u_2u_2}|,
\end{eqnarray}
where the specific value of the cut-off $\omega_H=k_B T_H /\hbar$ is determined by the analog of the surface gravity of the horizon configuration under examination~\cite{unruh}.

Before proceeding with the calculation of the density correlations, it is interesting
to point out a remarkable relation connecting the $S$-matrix of black hole and white hole configurations related by a time-reversal symmetry~\cite{macher2,macher1}:
\begin{equation}
\left[S^{WH}\right]^*=\left[S^{BH}\right]^{-1},
\eqname{S_BH_WH}
\end{equation}
with the convention that the $d$, $u_1$, $u_2$ modes of the white hole correspond after time-reversal (that exchanges the upstream and downstream regions) to the $u$, $d_1$, $d_2$ ones of the black hole as defined in~\cite{PRA_gradino}.
Making use of the $\eta$-unitarity of $S$, that is $S^\dagger \eta S =\eta$ with $\eta=\textrm{diag}[1,1,-1]$, this leads to
\begin{equation}
S^{BH}=\eta\,\left[S^{WH}\right]^T\,\eta.
\eqname{S_BH_WH2}
\end{equation}

\subsection{Density correlations between the internal and the external regions}
\label{sec:corr_inout}

The input-output approach reviewed in the previous section is the starting point for a calculation of the correlation function of density fluctuations. Such a calculation was first performed in~\cite{PRA_gradino} for the black hole case and is briefly reviewed in Sec.\ref{sec:anal_BH}. The results agreed with the numerical observations of~\cite{NJP_BH}. Here we shall extend this theory to the case of white holes.

Following the same lines as in~\cite{PRA_gradino}, one can show that the density correlation between the internal and the external regions of the white hole far away from the horizon can be written in the form
\begin{equation}
G^{(2)}(x,x')=1+\int_0^\infty\!d\omega\,\delta G_0^{(2)}(\omega,x,x')
\eqname{G2_an_gen}
\end{equation}
with
\begin{eqnarray}
n_0 \, \delta G^{(2)}_{0}(\omega,x,x') &=& \frac{1}{4\,\pi} \sum_{l=u_1,u_2}
\Big\{
\frac{R_{k^{\rm out}_d}(\omega)R_{k^{\rm out}_l}(\omega)}
{\sqrt{|v^{\rm out}_{g,d}\,v^{\rm out}_{g,l}|}}
\times
\nonumber
\\
&\times & e^{i [k^{\rm out}_d(\omega)  x'-k^{\rm out}_l(\omega)  x]}
\,\mathbf{S}^*_{l u_2}(\omega)\,\mathbf{S}_{d u_2}(\omega)+\rm{c.c.}\Big\}.
\eqname{G0inout}
\end{eqnarray}
Here, the point $x$ is taken in the upstream region inside the white hole, while the point $x'$ is taken in the downstream region outside the white hole. Of course, $\delta G^{(2)}_{0}(\omega,x,x')$ is non-zero only within the $\omega<\omega_{\rm max}$ region where the $u_2^{\mathrm in}$ mode exists. As already mentioned, we are here restricting our attention to the zero temperature case, $T_0=0$. 
Formula \eq{G0inout} clearly indicates that the process that is responsible for the correlation signal can be interpreted as the conversion of quantum fluctuations from the $u_2^{\rm in}$ mode into out-going pairs of $d^{\rm out}$ and $u_1^{\rm out}$ or $d^{\rm out}$ and $u_2^{\rm out}$ real Bogoliubov excitations: the $d^{\rm out}$ quantum is emitted away from the white hole in the rightwards direction, while the $u_{1,2}^{\rm out}$ quanta propagate upstream into the white hole.

We have defined $k_{d,u_1,u_2}^{\mathrm out}(\omega)$ as the value of the wavevector of the $d,u_1,u_2$ outgoing modes at the frequency $\omega$. The corresponding group velocities in the laboratory frame are defined as $v_{g,(d,u_1,u_2)}^{\rm out}$. The density components $R_{k_{d,u_1,u_2}^{\mathrm out}}(\omega)$ of the Bogoliubov modes are written in terms of the Bogoliubov $U_k,V_k$ coefficients as $R_k=U_k+V_k=[\hbar k^2/2m\Omega(k)]^{1/2}$~\cite{castin,BECbook}.
In particular, note how $R_{k_d^{\mathrm out}}(\omega)$ goes as $\sqrt{k_d^{\mathrm out}(\omega)}\sim \sqrt{\omega}$ in the low-frequency limit, while $R_{k_{u_{1,2}}^{\mathrm out}}(\omega)$ tend to finite values simply because both $k_{u_{1,2}}^{\mathrm out}(\omega\to 0)=\pm k_Z$: this simple fact will play a crucial role in the following.

\begin{figure}[htbp]
\begin{center}
\includegraphics[width=0.75\columnwidth,angle=0,clip]{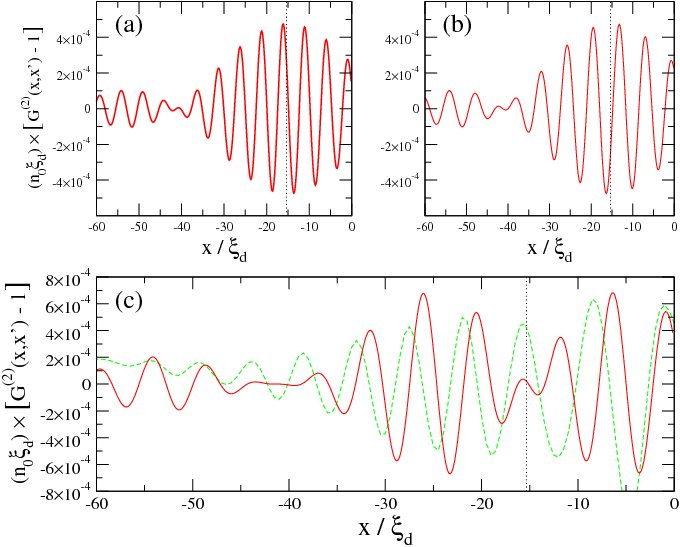}
\caption{Upper (a) and (b) panels, separated plots of the contribution of the $d$-$u_1$ and $d$-$u_2$ pairs to the correlation function of density fluctuations. Lower (c) plot, full analytical prediction \eq{final_inout} (red solid line). Numerical prediction evaluated at a late time $\mu_d t=180$ and rescaled to compensate for the spatial smoothening procedure (green dashed line). 
All the curves are cuts of the correlation function along the $x'=63.5\xi_d$ line (magenta line in Fig.\ref{fig:g2}). The vertical dotted line indicates the position of the intersection of the cut line with the destructive interference line \eq{center_inout}.
}
\label{fig:analytic_corr_inout}
\end{center}
\end{figure}

Approximating the ${S}$-matrix products with their low-energy form for a step-like horizon configuration
\begin{eqnarray}
S_{u_1u_2}^*S_{du_2}&=&\frac{1}{\sqrt{\omega}}
\frac{(v_0^2-c_u^2)^{3/4}(v_0-c_u)}{c_u(c_u-c_d)(c_d+v_0)}
\left[i\sqrt{v_0^2-c_u^2}-\sqrt{c_d^2-v_0^2}\right] \times \nonumber \\
&\times& \sqrt{\frac{c_d(c_d-v_0)}{2(c_d+v_0)
}} \\
S_{u_2u_2}^* S_{du_2}&=&-\left[S_{u_1u_2}^* S_{du_2}\right]^*,
\end{eqnarray}
and summing up the two terms, one is led to the final expression:
\begin{eqnarray}
n_0 \, \delta G^{(2)}_0(x,x')&\simeq& \frac{(v_0-c_u)\sqrt{(v_0^2-c_u^2)(c_d-v_0)}}{2\pi(c_d+v_0)^{5/2}c_u(c_u-c_d)} \times
\nonumber
 \\
&\times&\left[\sqrt{v_0^2-c_u^2} \, \cos(k_Z x)+\sqrt{c_d^2-v_0^2}\sin(k_Z x) \right] \times
\nonumber \\
&\times & \frac{1-\cos\left[\omega_{\rm max} \, \left( \frac{x}{v_Z}-\frac{x'}{v^{\rm out}_{g,d}} \right) \right]}{\frac{x}{v_Z}-\frac{x'}{v^{\rm out}_{g,d}}},
\eqname{final_inout}
\end{eqnarray}
where $v^{\rm out}_{g,d}=v_0+c_d$ is the speed of low-energy out-going $d$ phonons, while the low-frequency $u_{1,2}$ modes propagate at the zero-mode speed $v^{\rm out}_{g,u_{1,2}}\simeq v_Z$.

The carrier wavevector of the fringe system is determined by the finite wavevector $k_{u_{1,2}^{\rm out}}\simeq k_Z$ of the out-going $u_{1,2}^{\rm out}$ modes. In addition, its envelope shows a slower modulation at a wavevector $\omega_{\rm max}/v_Z$ set by the ultra-violet scale $\omega_{\rm max}$. The central minimum of the feature lies along the straight line
\begin{equation}
\frac{x}{v_Z}-\frac{x'}{v_0+c_d}=0
\eqname{center_inout}
\end{equation}
and is due to a destructive interference between the $u_{1,2}$ contributions.
This peculiar physics is illustrated in the cuts of the analytical pattern \eq{final_inout} that are shown in Fig.\ref{fig:analytic_corr_inout}: panels (a,b) show the separate contributions of the $u_1$ and $u_2$ modes: the carrier wavevector is set by $k_Z$, while the slow modulation is fixed by $\omega_{\rm max}$. The lower panel (c) shows their sum: destructive interference is responsible for the additional minimum in the envelope along the line \eq{center_inout}. Given the number of approximations involved in the analytical calculations leding to \eq{final_inout}, the qualitative agreement with the numerical results can be considered as quite good, in particular for what concerns the presence of the modulated envelope.

\subsection{Density correlations in the internal region: the growing checkerboard pattern}
\label{sec:corr_inin}

The same framework can be applied to the correlations between points located in the internal region of the white hole far away from the horizon.
In this case, auto-correlations in the $u_1$ and $u_2$ outgoing modes are important, as well as cross-correlations between the two modes.
Hence, the density correlation function is given by three contributions
\begin{eqnarray}
n_0\,\delta G_0^{(2)}(\omega,x,x')&=&\frac{1}{4\pi}\Big[
\frac{[R_{k^{\rm out}_{u_1}} 
]^2}
{|v^{\rm out}_{g,u_1}|} e^{i k^{\rm out}_{u_1}
\,(x'- x)}
\,|{S}_{u_1 u_2}
|^2+ \nonumber \\
&+&\frac{[R_{k^{\rm out}_{u_2}}
]^2}
{|v^{\rm out}_{g,u_2}|} e^{i k^{\rm out}_{u_2}
\,(x'- x)}
\,|{S}_{u_2 u_2}
|^2+ \nonumber \\
&+&\frac{R_{k^{\rm out}_{u_1}}
R_{k^{\rm out}_{u_2}}
}
{\sqrt{|v^{\rm out}_{g,u_1}\,v^{\rm out}_{g,u_2}|}}\,
e^{i [k^{\rm out}_{u_2}
x'-k^{\rm out}_{u_1}
x]}
\,{S}^*_{u_1 u_2}
\,{S}_{u_2 u_2}
+ \nonumber \\
&+&\rm{c.c.}\Big] +(x\leftrightarrow x').
\end{eqnarray}
The expressions for the low-frequency limit of the $S$-matrix elements within the step-like horizon approximation are
\begin{eqnarray}
S_{u_1u_2}^*S_{u_2u_2}&\simeq& \frac{(v_0^2-c_u^2)^{3/2}(c_d-v_0)}{2 \omega c_u(c_d+v_0)(c_u-c_d)^2}
\left[2v_0^2-c_d^2-c_u^2+2i\sqrt{(c_d^2-v_0^2)(v_0^2-c_u^2)} \right]\\
|S_{u_1u_2}|^2&\simeq& |S_{u_2u_2}|^2 \simeq \frac{(v_0-c_d)(v_0^2-c_u^2)^{3/2}(c_d+c_u)}{2c_u(c_d+v_0)(c_u-c_d)\omega}.
\end{eqnarray}
Inserting these expressions in \eq{G2_an_gen}, the integral over $\omega$ can be trivially performed. For the contribution of the $u^{\rm out}_1-u^{\rm out}_2$ cross-correlations and the sum of the $u^{\rm out}_{1}$ and $u^{\rm out}_2$ self-correlations, one obtains respectively:
\begin{eqnarray}
\left.n_0 \, \delta G_0^{(2)}(x,x')\right|_{u_1u_2}&\simeq& \frac{(v_0^2-c_u^2)(c_d-v_0)}{2\pi c_u(c_d+v_0)(c_u-c_d)^2}\times \nonumber \\
&\times& \Big\{ (2v_0^2-c_u^2-c_d^2)\cos[k_z(x+x')]+ \nonumber \\
&+&2\sqrt{(c_d^2-v_0^2)(v_0^2-c_u^2)}\sin[k_z(x+x')]\Big\}\,I_\epsilon, 
 \eqname{G2u1u2} \\
\left.n_0 \, \delta G_0^{(2)}(x,x')\right|_{u_1+u_2}&\simeq&
\frac{(v_0^2-c_u^2)(v_0-c_d)(c_d+c_u)}{2\pi c_u(c_d+v_0)(c_u-c_d)}\, \cos[k_z(x-x')]\,I_\epsilon. 
\eqname{G2u1u1u2u2}
\end{eqnarray}
The trigonometric factors in \eq{G2u1u2} and \eq{G2u1u1u2u2} are responsible for the checkerboard pattern that was observed in Fig.\ref{fig:g2}: the former gives fringes parallel to the anti-diagonal $x+x'=0$, while the former gives the fringes parallel to the main diagonal $x=x'$.
The most interesting factor is however the integral
\begin{equation}
I_\epsilon  =\int_\epsilon^{\omega_{\rm max}}\,\frac{d\omega}{\omega}\,\cos\left[\omega \left(\frac{x-x'}{v_Z}\right)\right],
\eqname{I}
\end{equation}
which diverges when the infrared cut-off $\epsilon \rightarrow 0$.

A way of coping with this divergence is to assume a scaling $\epsilon \propto 1/\tau$ for the infrared cut-off, $\tau$ being the time elapsed since the formation of the horizon.
In this way, $I_\epsilon$ becomes time-dependent: in particular, its value for $x=x'$ follows a logarithmic growth with $\tau$.
This fact is in agreement with the growth of the checkerboard pattern amplitude with time that was observed in the numerical simulations. In Fig.\ref{fig:cuts2}(a), we have tried to fit the numerical data with a logarithmic law (cyan line): the agreement is indeed quite good.

As a further check of our interpretation, we have repeated the same calculation for the finite temperature $T_0>0$ case, which introduces an additional $1/\omega$ factor into \eq{I}. The same strategy of imposing a infrared cut-off at $1/\tau$ then correctly explains the linear growth of the checkerboard pattern in time that was observed in the numerical calculation of Fig.\ref{fig:cuts2}(b). All these features will be discussed thoroughly in~\cite{Carlos_phd}.

\subsection{Analytic form of correlations in a black hole configuration}
\label{sec:anal_BH}

For the sake of completeness and to better understand the peculiarities of the white hole emission, it is useful to briefly review the analytical form of the different features that appear in the correlation pattern of black holes~\cite{PRA_gradino}.
The main ingredient of the calculation is again the $S$ matrix that connects the amplitude operators for the out-going $u^{\rm out},d_1^{\rm out},d_2^{\rm out}$ Bogoliubov modes to those of the in-going $u^{\rm in},d_1^{\rm in},d_2^{\rm in}$ ones.
From the relation \eq{S_BH_WH2}, it is immediately clear that for all $l=\{u,d_1,d_2\}$, $S_{ld_1}$ and $S_{ld_2}$ scale as $1/\sqrt{\omega}$, while $S_{lu}$ tends to a finite value in the low frequency limit.

\subsubsection{Density correlation between the internal and external regions}

The density correlation between the inside and outside regions of the black hole have the form:
\begin{eqnarray}
n_0 \, \delta G^{(2)}_{0}(\omega,x,x') &=& \frac{1}{4\,\pi} \sum_{l=d_1,d_2}
\Big\{
\frac{R_{k^{\rm out}_u}(\omega)R_{k^{\rm out}_l}(\omega)}
{\sqrt{|v^{\rm out}_{g,u}\,v^{\rm out}_{g,l}|}}
\times \nonumber \\
&\times&  e^{i [k^{\rm out}_u(\omega) x-k^{\rm out}_l(\omega) x']}
\,\mathbf{S}^*_{l d_2}(\omega)\,\mathbf{S}_{u d_2}(\omega)+\rm{c.c.}\Big\},
\eqname{BH_inout}
\end{eqnarray}
Here, the point $x>0$ is located in the upstream region outside the black hole, white the point $x'<0$ is located in the downstream region inside the black hole.

As in the white hole case, expression \eq{BH_inout} clearly indicates that the process responsible for the correlation signal consists of quantum fluctuations incident on the horizon from the $d_2$ mode, which scatter into a correlated pair of real out-going Bogoliubov excitations either into the $u$ and $d_1$ or into $u$ and $d_2$ modes. As already mentioned, in the black hole configuration the wavevector $k_l^{\rm out}$ of all out-going $l=\{u,d_1,d_2\}$ modes tends linearly to zero in the zero frequency limit, which implies that all $R_{k_l}^{\mathrm out}(\omega)\propto \sqrt{\omega}$.

For this reason, none of the black hole correlations shows any modulation and all features have a sharp tongue-like shape; within the low-frequency approximation of the $S$-matrix mentioned above, one indeed obtains
\begin{equation}
n_0\, \delta G^{(2)}_{0}(x,x')\simeq\!\!\!\sum_{l=d_1,d_2}\frac{A_l\,c_u^2\,\omega_{\rm max}\,
  \textrm{sinc}\left(\omega_{\rm max}(\frac{x}{v^{\rm out}_{g,u}}-\frac{x'}{v^{\rm out}_{g,l}})\right)}{4\pi
  |v^{\rm out}_{g,u} v^{\rm out}_{g,l} |\sqrt{c_u c_d}} \; ,
\label{eq:G2app}
\end{equation}
where $\textrm{sinc}(x)=\sin x /x$ and the $A_l$ coefficient has the following explicit form
\begin{equation}\label{eq:Al}
A_{d_1\;(d_2)}=\sqrt{\frac{c_u}{c_d}}
\frac{(c_u+ v_0)}{(c_u - v_0)}
\left(\frac{v_0^2}{c_u^2}-\frac{c_d^2}{c_u^2}\right)^{3/2}
\frac{c_u}{c_d+(-) c_u} \;
\end{equation}
in terms of the microscopic parameters of the system. $v^{\rm out}_{g,u}=v_0+c_u>0$ is the group velocity of the $u^{\rm out}$ mode escaping from the black hole. The group velocities $v^{\rm out}_{g,d_{1,2}}<0$ of the $d^{\rm out}_{1,2}$ outgoing modes are very different from each other, $v^{\rm out}_{g,d_{1,2}}=v_0\mp c_d$. This simple fact has the important consequence that the corresponding $u$-$d_2^{\rm out}$ and $u$-$d_1^{\rm out}$ features (iii) and (v) in the correlation pattern of Fig.\ref{fig:g2BH}(b) are spatially separated and do not interfere, as it was instead the case for a white hole case of Sec.\ref{sec:corr_inout}. Feature (iii) may be considered the main signature on the density correlation function of the Hawking emission from acoustic black holes.

\subsubsection{Density correlations in the internal regions}

The correlation function for points both located inside the black hole, $x,x'<0$ contains a single contribution that originates from the $d_1,d_2$ modes, namely
\begin{eqnarray}
n_0\, \delta G^{(2)}_{0}(\omega,x,x')&= &\frac{1}{4\,\pi}
\left(
\frac{R_{k^{\rm out}_{d_1}}(\omega)R_{k^{\rm out}_{d_2}}(\omega)}
{\sqrt{|v^{\rm out}_{g,d_1}v^{\rm out}_{g,d_2}|}}
e^{- i (k^{\rm out}_{d_1}(\omega) x-k^{\rm out}_{d_2}(\omega) x')}\times\right. \nonumber \\
&\times& \left.
\mathbf{S}^*_{d_1 d_2}(\omega)\,
\mathbf{S}_{d_2 d_2}(\omega)+\rm{c.c.}\right) +(x\leftrightarrow x').
\label{eq:G2relevantII}
\end{eqnarray}
Within the same low-frequency expansion approximation, one obtains a tongue-shaped contribution of the form
\begin{equation}
n_0 \, \delta G^{(2)}_{0}(x,x')\simeq
\frac{B\,\Omega\,c_u^2\,\textrm{sinc}\left(\Omega(\frac{x}{v^{\rm out}_{g,d_1}}-
\frac{x'}{v^{\rm out}_{g,d_2}})\right)}{4\pi v^{\rm out}_{g,d_1} v^{\rm out}_{g,d_2}\, c_d}+(x\leftrightarrow x')\,.
\label{eq:G2IIapp}
\end{equation}
with
\begin{equation}\label{B}
B=-\frac{c_u}{2c_d}
\frac{(c_u+ v_0)}{(c_u - v_0)}\left(\frac{v_0^2}{c_u^2}-\frac{c_d^2}{c_u^2}\right)^{3/2}
\; .
\end{equation}
that correctly reproduces feature (iv) of Fig.\ref{fig:g2BH}(b).
In particular, differently from the white hole case, no any infra-red divergence appears in the integral, which confirms the stationarity in time of the Hawking radiation pattern.

\vspace{0.5cm}

In summary, in this Section we have developed a Bogoliubov theory of quantum fluctuations in a white hole configuration within the step-like approximation. This theory provides analytical formulas for the physical observables that were considered in the numerical simulations reported in the previous Section. In contrast to the thermal character of the Hawking radiation from a black hole, the spectral distribution of the emitted radiation outside the white hole turns out to be flat. The steady growth of the density fluctuations in time is signalled by infrared divergences in the Bogoliubov calculation. By imposing a suitable cut-off to the integrals one is able to reproduce the numerically observed trend. Inclusion of the terms describing the back-action of fluctuations onto the condensate is presently in progress and is expected to shine light on the effect of quantum fluctuations on the stability of the acoustic white hole configuration.

\section{Conclusions}
\label{sec:conclu}

In this paper we have reported a comprehensive investigation of the physics of acoustic white hole configurations. We have focussed our attention on a simplest model that is amenable to analytical and numerical work. Using the Gross-Pitaevskii equation for the condensate dynamics, we have established the dynamical stability of the white hole at the mean-field level: no exponentially growing perturbation of the horizon appears when this is hit by an incident wavepacket of Bogoliubov phonons. A nonlinear back-action effect due to the perturbation of the horizon by the incident wavepacket is visible as a stationary pattern of Bogoliubov-\u Cerenkov waves by the perturbed horizon; the amplitude of this pattern scales as the square of the incident wavepacket one.

Quantum fluctuation are included within a truncated Wigner numerical approach; the physics of an infinitely long condensate is reproduced by imposing absorbing boundary conditions to fluctuations. Signatures of the white hole emission by the horizon are identified in the correlation function of density fluctuations.  
Differently from the Hawking emission from black holes, the white hole emission does not consist of low-energy excitations only, but involves high wavevector modes. As a result, the density correlation pattern shows fastly oscillating features on the scale of the healing length. Furthermore, some of these features are not stationary in time, but indefinitely grow with time with a logarithmic (linear) law for a zero (finite) initial temperature.
All the features found in the numerical observations are fully recovered and physically interpreted by an analytical calculation within a step-like approximation for the horizon. Interestingly, this theory predicts a flat spectral distribution of the radiation emitted outside the acoustic white hole: this is in strike contrast to the thermal character of Hawking radiation from black holes.

Besides its intrinsic interest from the point of view of quantum fluctuations in condensed matter systems, our results may open important avenues in gravitational physics and quantum field theory in curved space-times: in spite of being related to each other by a simple time reversal symmetry, the physical properties of the quantum emission from black and white holes turn out to be significantly different, starting from the thermal vs. flat frequency distribution of the emission. 
On one hand, Hawking radiation from black holes is concentrated in the low-$k$ hydrodynamical modes whose dispersion has a linear, relativistic-like behaviour; for this reason, Hawking radiation has turned out to be very robust against modifications of the ultraviolet behaviour of the dispersion relation. On the other hand, white hole radiation is intrinsecally not an hydrodynamical process: because of the large associated blue shift, it is in fact highly sensitive to the ultraviolet behaviour of the Bogoliubov dispersion relation.
On this basis, we expect that white hole radiation may then provide an even better magnifying glass to test the intimate microscopic structure of the quantum vacuum.

\section*{Acknowledgements}

We are indebted to N. Pavloff and M. Rinaldi for stimulating discussions.
A. F. wishes to thank Generalitat Valenciana for financial support. A.F. and C.M. are supported in part by MICINN grant FIS2008-06078-C03-02. C.M. is indebted to the Spanish MEC for a FPU fellowship. C.M. would like to thank the Universit\`a di Bologna and the BEC Center in Trento for their charming hospitality during the elaboration of this work.

\end{document}